\documentclass[useAMS,usenatbib,referee]{mn2e}
\usepackage {graphicx} 

\def\aap{A\&A\,  }
\def\aj{AJ  }
\def\apj{ApJ\,  }
\def\apjl{ApJ\,  }
\def\apss{Astrophysics and Space Science  }
\def\araa{ARA\&A  }

\def\mnras{MNRAS\,  }
\def\pasj{PASJ\,  }

\def\rmp{Rev. Mod. Phys.  }

\title
{
Evolution of superbubbles  in a 
self-gravitating disk.
}
\author    [L. Zaninetti] 
           {L. Zaninetti  \\ 
            Dipartimento di Fisica, \\
           Via Pietro Giuria 1               \\
           10125 Torino, Italy               \\
  email:  zaninetti@ph.unito.it
}
\date      {~~~~~~~~~~~~~~~~~~~~~~~~~~}
\date{to be inserted}

\pagerange{\pageref{firstpage}--\pageref{lastpage}}
\pubyear{2012}
\everymath     {\rm}
\everydisplay  {\rm}
\begin{document}

\maketitle

\label{firstpage}

\begin{abstract}
The expansion of a  superbubble is investigated
both analytically and numerically.
Our model 
implements the thin layer approximation
in a vertical profile of density as given 
by an isothermal self-gravitating disk.
A precise  comparison with the results of numerical 
hydro-dynamics  is given.
Analogies  are drawn with the Kompaneets
equation  that  includes 
the  quadratic hyperbolic-secant law 
in the list  of the plane-parallel 
stratified media.
An  astrophysical application  is made  
to  the  superbubble connected  
with the two  worms 46.4+5.5 and  39.7+5.7.
The effects of the rotation  of the galaxy on the simulated 
radius and on the velocity  
are  introduced.   
The worms 
with their strong   limb-brightening 
visible  on astronomical maps
are explained in the framework 
of image theory.
\end{abstract}

\begin {keywords}
ISM: bubbles, 
ISM: clouds,
Galaxy: disk,
galaxies: starburst 
\end{keywords}

\section{Introduction}

The superbubble (SB) plays a relevant role 
in astrophysics because
(i) it transports material from the galactic plane  up to 
a great galactic height, 
(ii) it can be a site for the acceleration
of cosmic rays, see 
\cite{Higdon2005,Higdon2006,Butt2008,Ferrand2010}.
SBs have been observed in various bands:
for  the H~I maps, see \cite{Oey2002}, 
for  the optical and HI observations, see \cite{Oey2002},
and for  the  X-ray  maps, see  
\cite{Chu2008,Rodriguez2011,Jaskot2011};
often the image
shows a strong   limb-brightening,   which indicates
that the emitting layer is thin.
The  worm is another 
observed feature that may, or may not, 
be associated with a wall of an SB.
Galactic worms were 
first identified as irregular, vertical columns of
atomic gas stretching from the galactic plane;
now, similar structures are found
in radio continuum and infrared maps,
see for example~\cite{Koo1992,English2000,Baek2008}.
The models  that  explain  the SB  as being due to 
the combined  explosions  of supernova in a cluster  of
massive stars will now be briefly reviewed.
The hydrodynamical approximation, 
with the inclusion of interstellar density gradients, 
can  produce  a blowout  into the galactic halo,
see  \cite{MacLow1989,Melioli2009}.
Expansion in the presence of magnetic fields
has  been implemented in various  magneto-hydrodynamic 
codes, see \cite{Tomisaka1992,Rafikov2000}.
In  semi-analytical calculations,
the thin layer approximation
can be the key to obtaining 
the expansion of
the SB: see,  for example,
\cite{McCray1987,McCray1987b,MacLow1988}.
The thin layer approximation allows of finding the equation 
of motion
for an expansion in the framework of momentum conservation,
see  \cite{Dyson1997}, when the density of the 
surrounding medium is constant.
The case of an expansion in a medium with 
variable density is more complex
and an exponential and a power law vertical profile 
have been analysed,
see  \cite{Zaninetti2010h}.
The exponential and vertical profiles in density  
do not correspond
to some physical process of equilibrium.
The case of an isothermal self-gravitating disk (ISD) 
is 
an equilibrium  vertical profile which can be 
coupled with momentum
conservation.
The models cited leave some questions
unanswered or only partially answered:
\begin {itemize}
\item  Is it possible to calibrate the vertical profile 
       of an ISD?
\item  Is it possible to deduce an analytical formula for 
        the temporal evolution of an SB in the presence 
        of a vertical profile density as given by an ISD?
\item Is it possible to  deduce numerical results for an SB when 
      the  expansion starts at a given galactic height?
\item What is  the influence of galactic rotation 
      on the temporal evolution of an SB?
\item Can we explain the worms as a particular effect using 
      image   theory   applied to SBs?
\end{itemize}
In order  to answer these questions,
Section \ref{sec_motion} 
reviews the standard equation for  momentum conservation 
in an advancing shell,
Section \ref{sec_asymmetry}  introduces
a vertical  profile   in the number of particles
as  given by an ISD 
which  models an aspherical expansion,
Section \ref{sec_applications}
applies the new  law  of motion
to  the SB associated with GW~46.4+5.5,
and Section \ref{sec_image}
contains detailed information
on how to build an image of a  SB 
as  seen from the equatorial plane 
as well  from the poles.

\section{The symmetrical thin layer approximation}

\label{sec_motion}

The thin layer approximation assumes that all the swept-up 
gas accumulates infinitely in a thin shell just after
the shock front.
The conservation of radial momentum requires that 
\begin{equation}
\frac{4}{3} \pi R^3 \rho \dot {R} = M_0
\quad,
\end{equation}
where $R$ and $\dot{R}$   are  the radius and the velocity
of the advancing shock,
$\rho$ the density of the ambient medium,
$M_0$ the momentum evaluated at $t=t_0$,
$R_0$ the initial radius,  
and 
$\dot {R_0}$  the  initial velocity,
see \cite{Dyson1997,Padmanabhan_II_2001}.
The law of motion is 
\begin{equation}
R = R_0 \left  ( 1 +4 \frac{\dot {R_0}} {R_0}(t-t_0) \right )^{\frac{1}{4}}  
\label{radiusm}
\quad ,
\end{equation}  
and the velocity 
\begin{equation}
\dot {R} = \dot {R_0} \left ( 1 +4 \frac{\dot {R_0}} {R_0}(t-t_0)\right )^{-\frac{3}{4}}  
\label{velocitym} 
\quad . 
\end{equation}

\section{Asymmetrical  law  of motion }

\label{sec_asymmetry}

Given the Cartesian   coordinate system
$(x,y,z)$,
the plane $z=0$ will be called the equatorial plane,
$z= R \sin ( \theta) $,
where $\theta$ is the latitude angle
which has range  
$[-90 ^{\circ}  \leftrightarrow  +90 ^{\circ} ]$,
and   $R$ is the distance from the origin.
The latitude angle is  often used  in
astrophysics  to model asymmetries in
the  polar lobes,
see the example of the nebula around
 $\eta$-Carinae (Homunculus)  shown in Table 1 in  
 \cite{Smith2002}.
In our  framework,  the polar angle of the spherical
coordinate system
is  $90 - \theta$.
The vertical number density distribution
of galactic H\,I  is usually modeled by  the following
three component  behavior as a function of
the  galactic height
{\it z}  which is the  distance  
from  the galactic plane in pc:
\begin{equation}
n(z)  =
n_1 e^{- z^2 /{H_1}^2}+
n_2 e^{- z^2 /{H_2}^2}+
n_3 e^{-  | z |  /{H_3}}
\,.
\label{exponential}
\end{equation}
We set the densities in Eq.~(\ref{exponential}) 
to $n_1 = 0.395$
particles cm$^{-3}$, $n_2 = 0.107$ particles cm$^{-3}$, $n_3 =
0.064$ particles cm$^{-3}$, and the scale heights to $H_1 = 127$
pc, $H_2 = 318$ pc, and $H_3 = 403$ pc.
(\cite{Lockman1984,Dickey1990,Bisnovatyi1995}). 
This  distribution
of  galactic H\,I is valid in the range 
0.4 $\leq$  $R$ $\leq$ $R_0$, 
where  $R_0$ = 8.5 kpc and $R$  is the distance
from  the galaxy centre.
Here, conversely,  we adopt 
the density profile of a thin
self-gravitating disk of gas which is characterized by a
Maxwellian distribution in velocity and  distribution which varies
only in the $z$-direction (ISD).
The  number density
distribution is  
\begin{equation}
n(z) = n_0 sech^2 (\frac{z}{2\,h})
\quad ,
\label{sech2}
\end{equation}
where $n_0$ is the density at $z=0$,
$h$ is a scaling parameter, 
and  $sech$ is the hyperbolic secant  
(\cite{Spitzer1942,Rohlfs1977,Bertin2000,Padmanabhan_III_2002}).

\begin{figure}
  \begin{center}
\includegraphics[width=6cm]{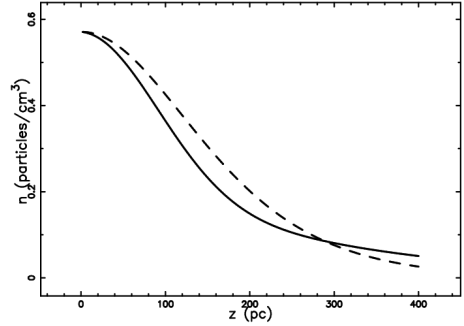}
  \end {center}
\caption
{
Profiles of density versus scale height $z$:
the  self-gravitating disk as
given by Eq.~(\ref{sech2})
when $h=90$\ pc
(dashed)
and
the
three-component exponential distribution
as
given by Eq.~(\ref{exponential})
(full line).
}%
    \label{f01}
    \end{figure}

Fig.~(\ref{f01})  compares 
the  empirical   function  sum of three exponential disks
and the theoretical  function
as given by Eq.~(\ref{sech2}).
Assuming that the expansion starts  at 
$z=0$, we can write $z=R \sin (\theta)$, 
and therefore
\begin{equation}
n(R,\theta) = n_0 sech^2 (\frac{R \sin (\theta) }{2\,h})
\quad  ,
\label{sech2rtheta}
\end{equation}
where  $R$ is the radius of the advancing shell.

The 3D expansion that starts at the origin 
of the coordinates 
will be characterized by the following
properties.
\begin {itemize}
\item The dependence of the momentary radius of the shell
      on  the latitude  angle $\theta$ over the range
      $[-90 ^{\circ}  \leftrightarrow  +90 ^{\circ} ]$.

\item The independence of the momentary radius of the shell
      from  $\phi$, the azimuthal  angle  in the $x$-$y$  plane,
      which has a range
      $[0 ^{\circ}  \leftrightarrow  360 ^{\circ} ]$.
\end {itemize}
The mass swept, $M$,  along the solid angle
$ \Delta\;\Omega $  between 0 and $R$ is
\begin{equation}
M(R,\theta)=
\frac { \Delta\;\Omega } {3}  m_H n_0 I_m(R)
+ \frac{4}{3} \pi R_0^3 n_0 m_H
\quad  ,
\end {equation}
where
\begin{equation}
I_m(R)  = \int_{R_0} ^R r^2 
sech^2 (\frac{r \sin (\theta) }{2\,h})
dr
\quad ,
\end{equation}
where $R_0$ is the initial radius
and $m_H$ the mass of hydrogen.
The integral is
\begin{eqnarray}
I_m(R)  =  
-4\,h{r}^{2} ( \sin ( \theta )  ) ^{-1} ( 1
+{{\rm e}^{{\frac {r\sin ( \theta ) }{h}}}} ) ^{-1}+4
\,{\frac {h{r}^{2}}{\sin ( \theta ) }}
\nonumber \\
-8\,{h}^{2}r\ln 
 ( 1+{{\rm e}^{{\frac {r\sin ( \theta ) }{h}}}}
 )  ( \sin ( \theta )  ) ^{-2}
-8\,{h}^{3}{
\it 
\mathrm{Li}_{2}\/}
(-{{\rm e}^{{\frac {r\sin ( \theta ) }
{h}}}} )  ( \sin ( \theta )  ) ^{-3}
\quad ,
\end{eqnarray}
where 
\begin{equation}
\mathop{\mathrm{Li}_{2}\/}\nolimits\!\left(z\right)=\sum
_{{n=1}}^{\infty}\frac{z^{n}}{n^{2}}
\quad ,
\end{equation}
is the  dilogarithm,
see  \cite{Hill1828,Lewin1981,NIST2010}.

The conservation of momentum 
along the solid angle
$ \Delta\;\Omega $
gives
\begin{equation}
M(R,\theta)     \dot {R} (\theta) =
M(R_0)          \dot {R_0}
\quad  ,
\end{equation}
where $\dot {R}(\theta)$  is the  velocity
at $R$ and
$\dot {R_0}$  is the  initial velocity at $R=R_0$.
Using the previous equation, 
an analytical  expression  
for  $\dot {R}(\theta)$
along the solid angle
can be found, 
but it is complicated, and therefore we omit it.
In this differential equation of the first order in $R$, the
variables can be separated and integrating
term by term gives
\begin{equation}
\int_{R_0}^{R}  M(r,\theta)  dr =
M(R_0) \dot {R_0} \, ( t-t_0)
\quad  ,
\end{equation}
where  
$t$ is the time and 
$t_0$ the time at $R_0$.
We therefore have  an equation of the type 
\begin{equation}
{\mathcal{F}}(R,R_0,h)_{NL} =
\frac{1}{3} R_0^3
\dot {R_0} \, 
\left( t-{\it t_0}
 \right) 
\quad  ,
\label{fundamental}
\end{equation}
where  $  {\mathcal{F}}(R,R_0,h)_{NL}  $ has an analytical
but complicated form.
The  case  of expansion  that starts  from a given 
galactic height $z$,  denoted by $z_{\mathrm{OB}}$,
which  represent  the OB associations,  
cannot be solved by Eq. (\ref{fundamental}),  which 
is derived  for  a  symmetrical expansion that 
starts at  $z=0$.
It is not possible to find  $R$   analytically  and
a numerical method   should be implemented.
In  our case, in order
to find  the root of the nonlinear
Eq. (\ref{fundamental}), 
the FORTRAN subroutine  ZRIDDR from \cite{press} has been used.

The following two recursive equations are found when
momentum conservation is applied:
\begin{eqnarray}
R_{n+1} = R_n + V_n \Delta t    \nonumber  \\
V_{n+1} = V_n 
\Bigl (\frac {M_n(r_n)}{M_{n+1} (R_{n+1})} \Bigr ) 
\quad  ,
\label{recursive}
\end{eqnarray}
where  $R_n$, $V_n$, $M_n$ are the temporary  radius,
the velocity,  and the total mass, respectively,
$\Delta t $ is the time step,  and $n$ is the index.
The advancing expansion is computed in a 3D Cartesian
coordinate system ($x,y,z$)  with the centre 
of the explosion at  (0,0,0).
The explosion is better visualized  
in a 3D Cartesian
coordinate system ($X,Y,Z$) in which the galactic plane
is given by $Z=0$.
The following 
translation, $T_{\mathrm{OB}}$,   
relates  the two Cartesian coordinate  systems. 
\begin{equation}
T_{\mathrm{OB}} ~
 \left\{ 
  \begin {array}{l} 
  X=x  \\\noalign{\medskip}
  Y=y  \\\noalign{\medskip}
  Z=z+ z_{\mathrm{OB}}
  \end {array} 
  \right.  \quad , 
\label{ttranslation}
\end{equation}
where $z_{\mathrm{OB}}$  
is the distance  in parsecs  of the 
OB associations   from the galactic plane.

The physical units have not yet been specified: 
parsecs for length and
$10^7\,yr$ for time are perhaps an acceptable 
astrophysical choice. 
With
these units, the initial velocity $V_{{0}}=\dot {R_0}$ is 
expressed in
units of pc/($10^7$ yr) and should be converted 
into km/s; this
means that $V_{{0}} =10.207 V_{{1}}$ 
where  $V_{{1}}$ is
the initial velocity expressed in km/s.

Analytical  results  can also be obtained 
solving  the Kompaneets
equation, see \cite{Kompaneets1960}, 
for the motion
of a shock wave in different plane-parallel 
stratified media   such as 
exponential, power-law type, and a 
quadratic hyperbolic-secant, 
see synoptic Table  4 in   \cite{Olano2009}.

\section{Astrophysical applications}

\label{sec_applications}

A  useful formula that allows setting up the initial 
conditions  is   formula (10.38)  
in  \cite{McCray1987} 
which models the SB  evolution 
in the energy conserving phase  when the 
number density is constant,  
\begin{equation}
R =111.56\;  \mathrm{pc}(\frac{E_{51}t_7^3 N^*}{n_0})^{\frac {1} {5}}
,
\label{raggioburst}
\end{equation}
where $t_7$ is the time 
 expressed  in units of $10^7$ yr,
$E_{51}$  is the  energy expressed  in  units of $10^{51}$ erg,   
$n_0$ is  the number density expressed  in 
particles~$\mathrm{cm}^{-3}$
(density~$\rho_0=n_0m$, where $m=1.4m_{\mathrm {H}}$) and
$N^*$  is the number of SN explosions
in  $5.0 \cdot 10^7$ yr and therefore is a rate,
see  \cite{McCray1987}.
This formula, deduced in spherical
coordinates, can be used only  
when  the density is constant, as  an example
for the first 20 pc,  the variation
of the density with galactic height
is $\approx~2\%$.  
In the following, we will  assume that 
the bursting phase  ends at $t=t_{7,0}$   (the bursting time is expressed in
units of $10^7$ yr) 
when  $N_{SN}$ SN are exploded 
\begin{equation}
N_{SN} = N^* \frac{t_{7,0} \cdot  10^7} {5 \cdot 10^7}
\quad .
\end{equation}   

The velocity of the SB   
in the energy conserving phase  when the number 
density is constant   is  
\begin{equation}
V(t_7) = \frac
{
6.567\,\sqrt [5]{{\frac {{\it E51}\,{\it N^*}}{{\it n_0}}}}
}
{
{{\it t_7}}^{2/5}
}
\frac {km}{s}
\quad .
\label{velburst}
\end{equation}
Eq. (\ref{raggioburst}) and (\ref{velburst})
can be used to deduce  $R_0$  and  
and  $V_0$  once  $N^*$  and  $t_{7,0}$ are given.
We continue giving  some information about the 
astrophysical  target of our simulation, 
analysing the analytical  solution at  
$z_{\mathrm{OB}}$ =0 and the numerical solution 
$z_{\mathrm{OB}} \neq 0 $,
making  a comparison with the  hydro code 
and  analysing the effects of the galactic rotation
on the obtained results.

\subsection{The simulated object}

When the  two  worms 46.4+5.5 and  39.7+5.7 
are carefully studied (\cite{Kim2000}) 
it is possible to conclude that 
they  belong to a single
SB. 
The parameters  of  
this  single  SB 
are given in  Table~\ref{tab:ssh}.

   \begin{table}
      \caption{Data of the SB associated with GW~46.4+5.5.}
         \label{tab:ssh}
      \[
         \begin{array}{cc}
            \hline
            \noalign{\smallskip}
\mbox {Dimensions~(pc}^2)                    & 345 \times 540  \\
\mbox {Expansion~velocity~ (km~s$^{-1}$}) & 15              \\
\mbox {Age~(10$^7$~yr})                    & 0.5             \\
\mbox {z$_{OB}$  (pc)}                   & 100             \\
\mbox {Total~energy~($10^{51}$}{\mathrm{erg}})      & 15              \\
            \noalign{\smallskip}
            \hline
         \end{array}
      \]
   \end{table}
 
\subsection{Analytical  versus numerical solutions}

\begin{figure}
  \begin{center}
\includegraphics[width=6cm]{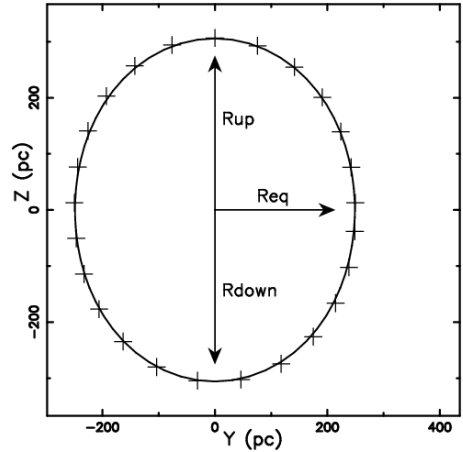}
  \end {center}
\caption
{
Section of the SB GW~46.4+5 in the {\it Y-Z;X=0}  
plane when the explosion starts at  $z_{\mathrm{OB}}=0 \mbox{pc} $.  
The analytical results  are shown as full 
line and the numerical 
results as crosses.  
The code parameters of the solution of 
Eq.~(\ref{fundamental}) as well of the  numerical couple  
(\ref{recursive})
are 
$h=90$ pc, 
$t_7=0.45~$,
$t_{7,0}$ = 0.0045, 
$r_0$ = 49.12, 
$V_0=641.7\, \mathrm{km}\,\mathrm{s}^{-1}$,
$N_{SN}$ = 93   
 and  $N^*$=103000.}
\label{f02}%
    \end{figure}

\begin{figure}
  \begin{center}
\includegraphics[width=6cm]{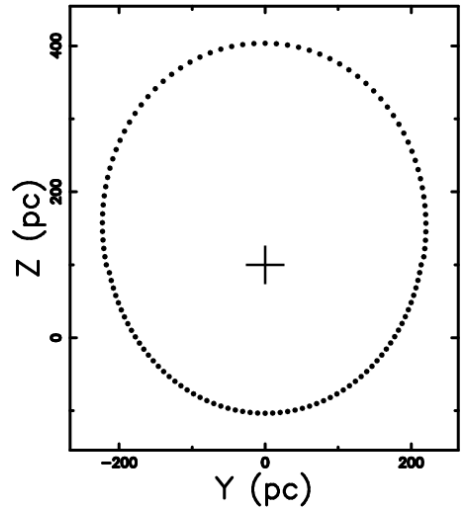}
  \end {center}
\caption
{
Section of the SB 
GW~46.4+5
in the {\it Y-Z;X=0}  plane
when the explosion starts at  $z_{\mathrm{OB}}=100 \mbox{pc}$.
The code parameters for  the  numerical couple  
(\ref{recursive})
are 
$h=90$ $pc$, 
$t_7=0.45~$,
$t_{7,0}$ = 0.00045, 
$r_0$ = 24.43, 
$V_0=3191 \mathrm{km}\, \mathrm{s}^{-1}$,
$N_{SN}$ = 180  
 and  $N^*$=2000000.
The explosion site is represented by a cross.
}
\label{f03}%
    \end{figure}

The analytical   model as  
given by the solution of the  nonlinear  
Eq. (\ref{fundamental}) can be used  only  in the
case $z_{\mathrm{OB}}$=0.
As an example,  we give a model of the 
SB associated with GW~46.4+5.5, 
see  Fig. \ref{f02}.
The  numerical  solution as given by 
the recursive relationship (\ref{recursive})
can be found adopting the same  input data
of  the analytical solution, 
see crosses  in  Fig. (\ref{f02});
the numerical solution agrees with the analytical 
solution within 0.65\%.
We  are now ready to present the numerical evolution 
of the SB   associated with GW~46.4+5
when $z_{\mathrm{OB}}=100 \mbox{pc} $,
see Fig.~\ref{f03} and
Table \ref{tab:ssh:simulated}.
   \begin{table}
      \caption{Simulated data of the SB 
         associated with GW~46.4+5.5.}
         \label{tab:ssh:simulated}
      \[
         \begin{array}{cc}
            \hline
            \noalign{\smallskip}
\mbox {Size~(pc}^2)                    & 454 \cdot 521  \\
\mbox {Averaged expansion~velocity~ (km~s$^{-1}$}) & 19.3              \\
\mbox {Age~(10$^7$~yr})                    & 0.45             \\
\mbox {z$_{OB}$  (pc)}                   & 100             \\
            \noalign{\smallskip}
            \hline
         \end{array}
      \]
   \end{table}

\subsection{Numerical solution and hydro code}

\begin{figure}
  \begin{center}
\includegraphics[width=6cm]{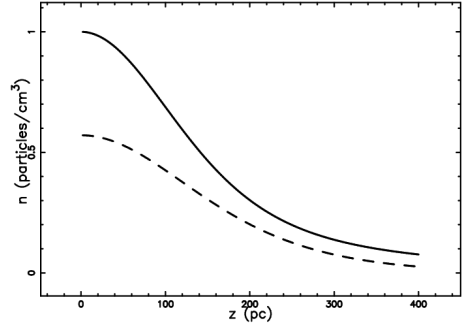}
  \end {center}
\caption
{
Profiles of density versus scale height $z$:
the  self-gravitating disk as
in Eq.~(\ref{sech2})
when $h=90$\ pc
(dashed line)
and the hydro number density as given by Eq. \ref{ISMhydro}
(full line).
}%
    \label{f04}
    \end{figure}

The level  of confidence in   our results
can  be given by a comparison with numerical 
hydro-dynamics 
calculations (see, for example, \cite{MacLow1989}).
The vertical density distribution
they  adopted,  see equation (1) in \cite{MacLow1989} and
equation (5) in~\citet{Tomisaka1986},
has  the following dependence on $z$, 
the  distance  from  the galactic plane in parsec:
\begin {equation}
n_{\mathrm {hydro}} =
n_{\mathrm {d}} \left \{
\Theta {\mathrm {exp}} [-\frac {V_p(z)}{\sigma_{{\mathrm {IC}}}^2} ]
+(1-\Theta) {\mathrm {exp}}
\left  [-\frac {V_p(z)}{\sigma_{{\mathrm {{\mathrm {C}}}}}^2} \right ]
\right\}
,
\label{ISMhydro}
\end {equation}
with the gravitational potential
\begin{equation}
V_p(z)=
68.6 {\mathrm {ln}} \left 
[1 + 0.9565~{\mathrm { sinh}}^2 \left (0.758 \frac {z}{z_0}\right) \right ]
({\mathrm {km s}}^{-1})^2
.
\end {equation}
Here, $n{\mathrm {_d}}=1~\mbox{particles~cm}^{-3}$, 
$\Theta=0.22$,
$\sigma_{IC}=14.4{\mathrm { km~s}}^{-1}$,
$\sigma_{C}=7.1  {\mathrm { km~s}}^{-1}$,
and $z_0 =124~{\mathrm {pc}}$.
Fig.~(\ref{f04})  compares
the  hydro number density as given by Eq. \ref{ISMhydro}
and the theoretical  function
as given by  Eq.~(\ref{sech2}).
The difference in the density profiles from hydrosimulations and 
from the simple model adopted in this paper
are due to the fact  that the density 
at  $z=0$ is  assumed to be   
$n_{\mathrm {hydro}}=1~\mbox{particles~cm}^{-3} $
in the hydrosimulations, see \cite{MacLow1989}.
In our  model conversely  at  $z=0$ we have  
$n=0.566~\mbox{particles~cm}^{-3} $
as  in \cite{Lockman1984}.

\begin{figure}
\includegraphics[width=6cm]{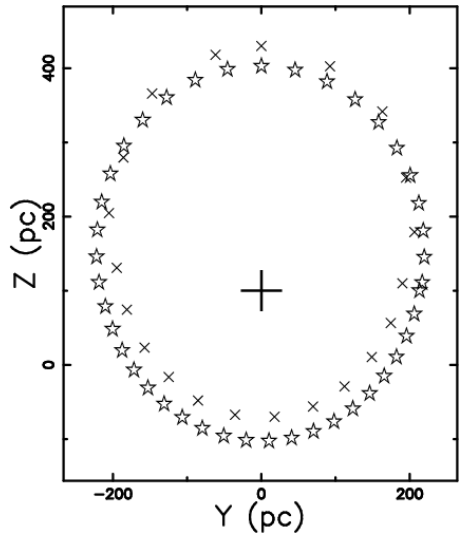}
\caption
{
Section of the SB in the {\it X-Z}  plane
when the explosion starts at  $z_{\mathrm{OB}}=100 \mbox{pc} $
(empty stars).
The code parameters are
$h=90$ pc, 
$t_7=0.45~$,
$t_{7,0}$ = 0.00045,
$r_0$  = 24.43, 
$V_0= 3191\, \mathrm{km}\, \mathrm{s}^{-1}$,
$N_{SN}$ = 180   
 and  $N^*$=2000000.
The points represented by the  small crosses indicate 
the inner section from  Fig. 3a  of  MacLow et al. 1989.
The explosion site is represented by a big cross.
}
\label{f05}%
    \end{figure}

A typical run  of ZEUS
(see~\cite{MacLow1989}),  a two-dimensional hydrodynamic   code, 
is  done for 
$t_7 =0.45 $ and   
 supernova luminosity of 
$1.6 \cdot 10^{38} {\mathrm{erg~s}}^{-1}$
when  $z_{\mathrm{OB}}$=100~pc.
In order to make a comparison with  our
code, we  adopt the same time  
and we search the  parameters 
which  produce similar  results, see  
Fig. \ref{f05}.
The percentage  of
reliability  of our code can 
also be  introduced,
\begin{equation}
\epsilon  =(1- \frac{\vert( R_{\mathrm {hydro}}- R_{\mathrm {num}}) \vert}
{R_{\mathrm {hydro}}}) \cdot 100
\,,
\label{efficiency}
\end{equation}
where $R_{\mathrm {hydro}}$ is the radius
given by the hydro-dynamics,
and  $R_{\mathrm {num}}$ s the radius  obtained from 
our  simulation.
Table~\ref{tab:rel}  shows 
our numerical radii 
in the    upward, downward, and  equatorial
directions,
and the efficiency  as  given by
formula~(\ref{efficiency}).

\begin{table}
      \caption{Code reliability. }
         \label{tab:rel}
      \[
         \begin{array}{ccccc}
            \hline
            \noalign{\smallskip}
~~~~     & R_{\mathrm{up}}(\mathrm{pc})  &
         R_{\mathrm{down}}(\mathrm{pc})  &
         R_{\mathrm{eq}}  (\mathrm{pc})  \\
            \noalign{\smallskip}
            \hline
            \noalign{\smallskip}
R_{\mathrm {hydro}} (ZEUS)            &  330  &  176   &  198        \\
R_{\mathrm {num}}   (\mbox{our~code}) &  302  &  203   &  217 
        \\
\mbox {efficiency} (\%)               &  91   &  84  &   90        \\
            \noalign{\smallskip}
            \hline
         \end{array}
      \]
   \end{table}

\subsection{Galactic rotation}

\begin{figure}
\includegraphics[width=6cm]{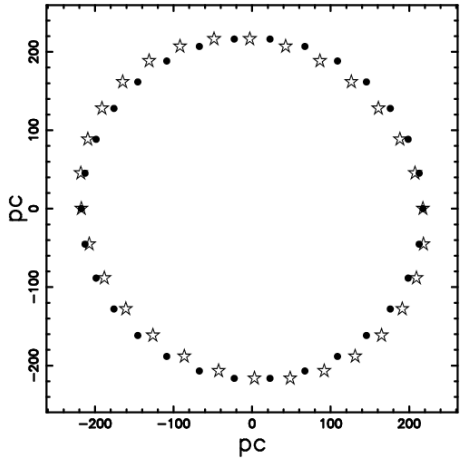}
\caption
{
Section of the SB 
GW~46.4+5
in the {\it Y-X,Z=0}  plane
when the explosion starts at  
$z_{\mathrm{OB}}=100 \mbox{pc} $.
Full points represent the circular  section,
the empty stars 
the rotation-distorted section.
Parameters as in Fig.~\ref{f03}
and $R_0[\mathrm{pc}]$=8500.
}
\label{f06}%
\end{figure}

\begin{figure}
\includegraphics[width=6cm]{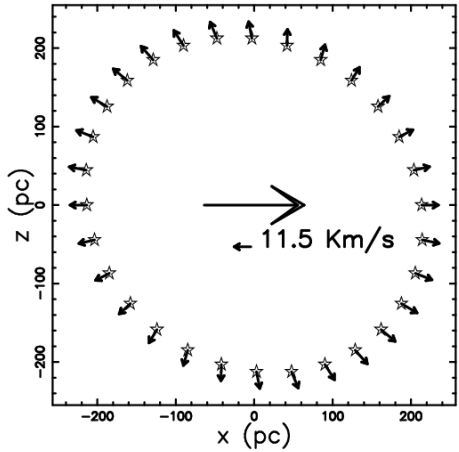}
\caption
{
The stars  represent  
the rotation-distorted section 
when $R_0[\mathrm{pc}]$=8500.
The arrows show the velocity field   of  
the expansion  as modified 
by the shear velocity.
The direction of  rotation of the Galaxy  
(big arrow) and the scale of the velocity are shown.  
}
\label{f07}%
\end{figure}

The  influence of the Galactic rotation on the results 
can be be obtained by  introducing 
the  law  of the Galactic rotation 
as given by \cite{Wouterloot_1990},
\begin{equation}
V_{\mathrm{R}} (R_0)  =220 ( \frac {R_0[\mathrm{pc}]} {8500})^{0.382} 
\mathrm {km~sec}^{-1}
,
\label {vrotation}
\end {equation} 
where  $R_0$ is the radial distance from 
the centre of the Galaxy 
in parsecs.
The original circular  shape of the superbubble  at a 
given value  of $z$ 
transforms to an ellipse by  the 
 following 
transformation, $T_{\mathrm {r}}$,   
\begin{equation}
T_{\mathrm{r}} ~
 \left\{ 
  \begin {array}{l} 
  x\prime=x  +  0.264 y ~t\\\noalign{\medskip}
  y\prime=y\\\noalign{\medskip}
  z\prime=z,
  \end {array} 
  \right. 
\label{trotation}
\end{equation}
where {\it y} is in parsecs 
and $t$ is expressed in units of $10^7~{\mathrm{yr}}$ 
\cite{Zaninetti2004}, 
see Fig. \ref{f06}.
In the same way, the effect of the shear velocity 
as a function of the distance  $y$  
from the centre of the expansion,
$V_{shift}(y)$,  
can be easily obtained by performing a Taylor expansion  
of Eq.~(\ref{vrotation})
\begin{equation}
V_{shift}(y) = 84.04  \frac {y}{R_0}
\, Km/s 
\quad .
\label{vshift}
\end{equation}
This is the the shear velocity as a function
of $y$.  Fig. \ref{f07} shows
the rotation-distorted map of  position and velocity.

\begin{figure}
\includegraphics[width=6cm]{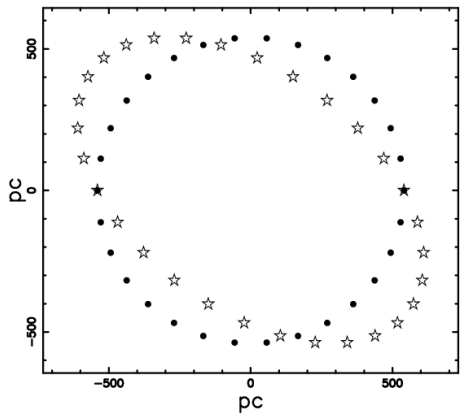}
\caption
{
Section of the SB 
GW~46.4+5 in the {\it Y-X,Z=0}  plane
when the explosion starts at  
$z_{\mathrm{OB}}=100 \mbox{pc}$.
Full points represent the circular  section and 
the empty stars  the rotation-distorted section.
Parameters as in Fig.~\ref{f03}
except  $t_7=2~$ 
and $R_0[\mathrm{pc}]$=8500.
}
\label{f08}%
\end{figure}

The effect of rotation  seems to be small
and  in order to see  a  more relevant correction
we should  increase  the  elapsed time,
see  Fig. \ref{f08}.    

\section{The image }

\label{sec_image}
The transfer equation in the presence of emission
only, see for
example \cite{rybicki} or \cite{Hjellming1988}, is
 \begin{equation}
\frac {dI_{\nu}}{ds} =  -k_{\nu} \zeta I_{\nu}  + j_{\nu} \zeta
\label{equazionetrasfer} \quad,
\end {equation}
where  $I_{\nu}$ is the specific intensity, 
$s$  is the line of sight, 
$j_{\nu}$ the emission coefficient, $k_{\nu}$   a mass
absorption coefficient, $\zeta$ the  mass density
at position $s$,
and the index $\nu$ denotes the relevant
frequency of emission.
The solution to  Eq.~(\ref{equazionetrasfer})
 is
\begin{equation}
 I_{\nu} (\tau_{\nu}) =
\frac {j_{\nu}}{k_{\nu}} ( 1 - e ^{-\tau_{\nu}(s)} ) \quad ,
\label{eqn_transfer}
\end {equation}
where $\tau_{\nu}$ is the optical depth 
at frequency $\nu$
\begin{equation}
d \tau_{\nu} = k_{\nu} \zeta ds \quad.
\end {equation}
We now continue analysing the case of an
optically thin layer
in which $\tau_{\nu}$ is very small
( or $k_{\nu}$  very small )
and the density  $\zeta$ is replaced with our number density
$C(s)$ of  particles.
One case is taken into account:   the
emissivity is proportional to the number density
\begin{equation}
j_{\nu} \zeta =K  C(s) \quad ,
\end{equation}
where $K$ is a  constant.
This  can be the case for synchrotron emission 
and   
a simple model  for the acceleration of the electrons
can be found in  Appendix \ref{appendice}.
We select  
as an example the  [S II] continuum  of 
the synchrotron superbubble
in the irregular galaxy
IC10, see \cite{Lozinskaya2008},
 and the 
X-ray emission below 2 keV  
around the OB association LH9 
in the H II complex N11 in the Large
Magellanic Cloud, see \cite{Maddox2009}.
The intensity at 
a given frequency 
is 
\begin{equation}
I (\nu) \propto  l  \nu^{\beta }
\quad, 
\end {equation} 
where $l$ is the length of the radiating region 
along the line of sight.
The source of synchrotron luminosity
is assumed here to be
the rate of kinetic energy,
$L_m$,
\begin{equation}
L_m = \frac{1}{2}\rho A  V^3
\quad,
\label{fluxkineticenergy}
\end{equation}
where $A$ is the considered area, see formula (A28)
in \cite{deyoung}.
In the case of formula (\ref{raggioburst})   
$A=4\pi R^2$,
which means
\begin{equation}
L_m = \frac{1}{2}\rho 4\pi R^2 V^3
\quad,
\label{fluxkinetic}
\end{equation}
where $R$  is the instantaneous radius of the SB  and
$\rho$  is the density in the advancing layer
in which the synchrotron emission takes place.
The astrophysical  version of the 
the rate of kinetic energy,
\begin{equation}
L_{ma} =
{ 1.39\times 10^{29}}\,{\it n_1}\,{{\it R_1}}^{2}{{\it V_1}}^{3}
\frac{ergs}{s}
\quad,
\label{kineticfluxastro}
\end{equation}
where $n_1$   is the   number density expressed
in units  of  $1~\frac{particle}{cm^3}$,
$R_1$  is  the  radius in parsecs,
and
$V_{1}$ is the   velocity in
km/s.
The spectral luminosity, $ L_{\nu} $,
at a given frequency $\nu$
is
\begin{equation}
L_{\nu} =  4 \pi  D^2  S_{\nu}
\quad ,
\end{equation}
with
\begin{equation}
S_{\nu} =  S_
0  (\frac{\nu}{\nu_0})^{\beta}
\quad ,
\end{equation}
where  $S_0$   is the flux
observed at  the frequency
$\nu_0$  and  $D$ is the  distance.
The total observed synchrotron 
luminosity, $L_{tot} $,
is
\begin{equation}
L_{tot} =
\int_{\nu_{min}}^{\nu_{max}}  L_{\nu} d \nu
\quad ,
\end{equation}
where
${\nu_{min}}$ and
${\nu_{max}}$
are the  minimum and maximum frequencies  observed.
The  total observed
luminosity
can  be expressed as
\begin{equation}
L_{tot} = \epsilon  L_{ma}
\label{luminosity}
\quad ,
\end{equation}
where  $\epsilon$  is  a constant  of conversion
from  the mechanical luminosity   to  the
total observed luminosity in synchrotron emission.

\begin{figure*}
\begin{center}
\includegraphics[width=6cm]{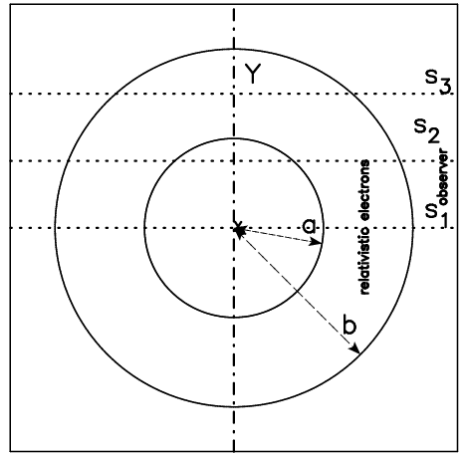}
\end {center}
\caption
{
The two circles (sections of spheres)  which   
include the region
with constant density
are   represented through
 a full line.
The observer is situated along the $x$ direction,  
three lines of sight are indicated,
and  the relativistic  electrons have  radius $r$  
in the region   $a <r <b$.
}
\label{f09}
    \end{figure*}

The fraction  of the total  luminosity deposited  in a
band  $f_c$  is
\begin{equation}
f_c  =
\frac
{
{{\it \nu_{c,min}}}^{\beta+1}-{{\it \nu_{c,max}}}^{\beta+1}
}
{
{{\it \nu_{min}}}^{\beta+1}-{{\it \nu_{max}}}^{\beta+1}
}
\quad ,
\end{equation}
where  $\nu_{c,min}$  and  $\nu_{c,max}$
are the minimum and maximum frequency  of the band.
Table \ref{tablecolors} shows some values of 
$f_c$  for the most important optical bands.
\begin{table} [h!]
\label{tablecolors}
\begin{center}
      \caption
      {
      Table of the values of $f_c$ when
      $\nu_{min}= 10^7 Hz$,
      $\nu_{max}= 10^{18} Hz$
      and  $\beta=-0.7$.
         }
         \begin{tabular}{crrr}
            \hline
           \hline
band         & $\lambda$  ($\AA$) &  FWHM  ($\AA$) & $f_c$\\
            \hline
U            &  3650   &  700  & 6.86 $\times  10^{-3}$  \\
B            &  4400   &  1000 & 7.70 $\times  10^{-3}$   \\
V            &  5500   &  900  & 5.17 $\times  10^{-3}$  \\
$ H\alpha$   &  6563   &  100  & 0.56 $\times  10^{-3}$  \\
$[S II]$ continuum   &  7040   &  210  & 0.92 $\times  10^{-3}$  \\
            \hline
            \hline
         \end{tabular}
   \end{center}
   \end{table}
An analytical  solution  for the radial cut of intensity 
of  emission can be found  in the equatorial plane 
$z_{\mathrm{OB}}$=0.
We assume that the number density 
of relativistic electrons 
$C$ is constant and in particular
rises from 0 at $r=a$ to a maximum value $C_m$, remains
constant up  to $r=b$, and then falls again to 0,
see \cite{Zaninetti2009a}.
This geometrical  description is shown in  
Fig.~\ref{f09}.
The length of sight, when the observer is situated
at the infinity of the $x$-axis, 
is the locus    
parallel to the $x$-axis which  crosses  the position $y$ 
in a 
Cartesian $x-y$ plane and 
terminates at the external circle
of radius $b$.
When the number density
of the relativistic electrons 
$C_m$ is constant between the two spheres
of radii $a$ and $b$, 
the intensity of radiation is 
\begin{eqnarray}
I_{0a} =C_m \times 2 \times ( \sqrt { b^2 -y^2} - \sqrt {a^2 -y^2}) 
\quad  ;   0 \leq y < a  \nonumber  \\
I_{ab} =C_m \times  2 \times ( \sqrt { b^2 -y^2})  
 \quad  ;  a \leq y < b    \quad . 
\label{irim}
\end{eqnarray}
The ratio between the theoretical intensity 
 at the maximum      ($(y=a)$)
 and at the minimum  ($y=0$)
is given by 
\begin{equation}
\frac {I(y=a)} {I(y=0)} = \frac {\sqrt {b^2 -a^2}} {b-a}
\quad .
\label{ratioteorrim}
\end{equation}

\begin{figure*}
\begin{center}
\includegraphics[width=6cm]{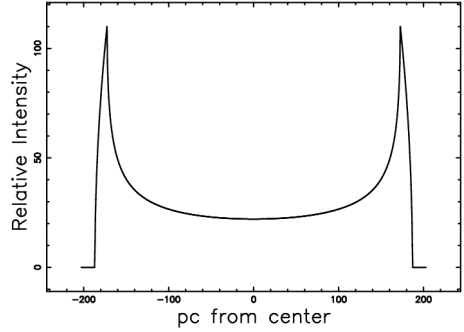}
\end {center}
\caption
{
 Cut of the mathematical  intensity ${\it I}$
 of the ring  model, Eq.~(\ref{irim}), 
 in the equatorial plane    
 (full  line) 
of the SB associated with GW~46.4+5.5~.
The $x$- and  $y$-axes  are in pc,
 $a=172.5 $ pc,  $b=186.9 $ pc
 and $\frac {I(y=a)} {I(y=0)}=5$.
}
\label{f10}
    \end{figure*}

\begin{figure}
\includegraphics[width=6cm]{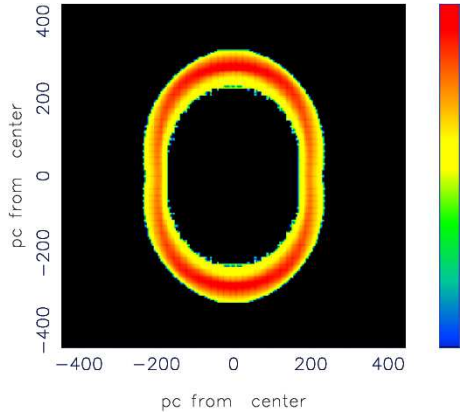}
\caption {
Map of the theoretical intensity  of
SB GW~46.4+5.
Physical parameters as in Fig. \ref{f02}.
The three Euler angles
characterizing the   orientation
  are $ \Phi  $=90 $^{\circ }$,
      $ \Theta$=90 $^{\circ }$
and   $ \Psi  $=90 $^{\circ }$.
This  combination of Euler angles corresponds
to the rotated image with the polar axis along the
z-axis.
In this map $I_{tr}= I_{max}/2 .$
}%
    \label{f11}
    \end{figure}

\begin{figure}
\includegraphics[width=6cm]{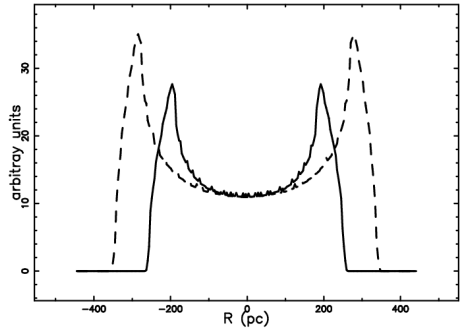}
\caption 
{
 Two cuts of the model intensity
 across the centre of the SB:
 equatorial cut (full line)
 and polar cut  (dotted line).
 Parameters as in Fig. ~\ref{f11} and
 $I_{tr}= 0$.
}
    \label{f12}
    \end{figure}

A cut in the theoretical intensity 
of the SB associated with GW~46.4+5.5 
is shown in Fig.~\ref{f10}.
Similar analytical results for the 
intensity of the cuts 
in the H$\alpha$  of planetary nebulae
and in the radio of supernova remnants 
have  been found
by \cite{Gray2012},
compare their Figure 5 with our
Figure \ref{f09} 
 and by \cite{Opsenica2011}, see their Figure 1.
A simulated image of the complex shape of a   SB 
is composed
by combining the intensities
which  characterize
different points of the advancing shell.
For an optically thin medium, the transfer equation
provides the emissivity to be multiplied with the distance
on  the line of sight, $l$.
This   length  in  SB depends
on the  orientation of the observer
but for the sake of clarity the observer is at infinity
and sees  the SB from the equatorial plane
$z_{\mathrm{OB}}$ =0
or from one of the two poles.  
We now outline 
the numerical algorithm which allows us to
build  the  complex  image  of an SB.
\begin{itemize}
\item 
An empty (value=0)
memory grid  ${\mathcal {M}} (i,j,k)$ which  contains
$NDIM^3$ pixels is considered
\item 
We  first  generate an
internal 3D surface by rotating the section of 
 $180^{\circ}$
around the polar direction and 
a second  external  surface at a
fixed distance $\Delta R$ from the first surface. 
As an example,
we fixed $\Delta R$ = $ 0.03 R_{max}$, 
where $R_{max}$ is the
maximum radius of expansion.
The points on
the memory grid which lie between the internal and the external
surfaces are memorized on
${\mathcal {M}} (i,j,k)$ with a variable integer
number   according to formula
(\ref{fluxkinetic})  and   density $\rho$ proportional
to the swept    mass.
\item Each point of
${\mathcal {M}} (i,j,k)$  has spatial coordinates $x,y,z$ 
which  can be
represented by the following $1 \times 3$  matrix, $A$,
\begin{equation}
A=
 \left[ \begin {array}{c} x \\\noalign{\medskip}y\\\noalign{\medskip}{
\it z}\end {array} \right]
\quad  .
\end{equation}
The orientation  of the object is characterized by
 the
Euler angles $(\Phi, \Theta, \Psi)$
and  therefore  by a total
$3 \times 3$  rotation matrix,
$E$, see \cite{Goldstein2002}.
The matrix point  is
represented by the following $1 \times 3$  matrix, $B$,
\begin{equation}
B = E \cdot A
\quad .
\end{equation}
\item
The intensity map is obtained by summing the points of the
rotated images
along a particular direction.
\item
The effect of the  insertion of a threshold intensity, $I_{tr}$,
given by the observational techniques,
is now analysed.
The threshold intensity can be
parametrized by $I_{max}$,
the maximum  value  of intensity
which  characterizes  the map,
see  \cite{Zaninetti2012b}. 
\end{itemize}
An  ideal image of the intensity of  SB GW~46.4+5  
is shown in Fig. \ref{f11}, and 
Fig. \ref{f12} shows two
cuts through the centre of the SB.

\begin{figure}
\includegraphics[width=6cm]{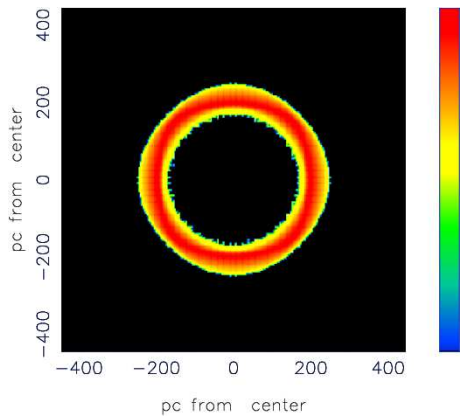}
\caption {
Map of the theoretical intensity  of
SB GW~46.4+5 as seen from the polar direction 
at infinity.
Physical parameters as in Fig. \ref{f02}.
The three Euler angles
characterizing the   orientation
  are $ \Phi  $=0$^{\circ }$,
      $ \Theta$=0$^{\circ }$
and   $ \Psi  $=0$^{\circ }$.
In this map $I_{tr}= I_{max}/2 .$
}%
    \label{f13}
    \end{figure}

We can also build the theoretical image as 
seen from one of the two
poles, see Fig.~\ref{f13}.

\section{Conclusions}

{\bf Law of motion}
The temporal evolution  of  an SB  in a medium 
with constant  density is  characterized  by a
spherical symmetry.
The presence  of a thin
self-gravitating disk of gas which is characterized by a
Maxwellian distribution
in velocity and  distribution of density which varies
only in the z-direction    produces 
an axial  symmetry in the temporal evolution 
of an SB.
The resulting Eq. (\ref{fundamental})  
has an 
analytical form which can be solved numerically
when $z_{\mathrm{OB}}$ =0.
The case of  $z_{\mathrm{OB}}\neq 0$ can be attached 
solving    two recursive equations, see (\ref{recursive}).
These complex shapes can  also be  modeled by 
the Kompaneets
equation when   a quadratic hyperbolic-secant
is adopted.  
Both  the  approximation here used and the 
Kompaneets equation can model 
the shape of the SB without
invoking the collision of two wind-blown SBs, see 
\cite{Ntormousi2011}.
The code developed  here runs in less than a minute on 
a LINUX -$2.66$GHz processor and can 
be an alternative to a purely numerical 2D or 3D model. 

{\bf Synchrotron Emission} 

Here we assumed that the conversion from 
the flux of kinetic energy into  non thermal luminosity
is regulated by a constant.
The exact value of this constant can be deduced once
the number density, radius, thickness, and    
velocity  of the advancing shell are provided by 
observational astronomy.
The values
of the fraction of flux deposited 
in the various  astronomical  bands, $f_c$,  are  given in 
Table \ref {tablecolors}.

{\bf Images}

The emissivity in the thin  advancing  layer is assumed
to be  proportional  to the  rate of  kinetic energy,
see  Eq.~(\ref{fluxkineticenergy}),
where  the density  is assumed to be
proportional  to the swept material.
This  assumption explains the strong   limb-brightening 
visible  on the astronomical maps 
of SBs and allows associating the observed 
worms to the non detectable SBs.
As an example, if the threshold of the observable 
flux at a given 
wavelength  is bigger  than the emitted flux  at the centre of 
an SB,  we  will  detect only the external regions, 
the so called worms, 
see  Fig. \ref{f12}.
\section*{ Acknowledgements}
I thank the  referee , Breitschwerdt Dieter  ,for constructive  
comments on the text.


\appendix 

\section {How to accelerate electrons}
\label{appendice}
An electron which  loses
its  energy  due to
synchrotron radiation
has a lifetime  of
\begin{equation}
\tau_r  \approx  \frac{E}{P_r} \approx  500  E^{-1} H^{-2} sec
\quad ,
\label {taur}
\end{equation}
where
$E$  is the energy in ergs,
$H$ the magnetic field in Gauss,
and
$P_r$  is the total radiated
power \citep[] [Eq. 1.157] {lang}.
The  energy  is connected  to  the critical
frequency \citep[][Eq. 1.154]{lang},
by
\begin {equation}
\nu_c = 6.266 \times 10^{18} H E^2~\mathrm{Hz}
\quad  .
\label {nucritical}
\end{equation}
The lifetime
for synchrotron  losses is
\begin{equation}
\tau_{syn} =
 39660\,{\frac {1}{H\sqrt {H\nu}}} \, \mathrm{yr}
\quad  .
\end{equation}

Following \citep{Fermi49,Fermi54},
the gain  in  energy  in a continuous    form
for  a particle
which  spirals  around a line of force
is  proportional to its
energy, $E$,
\begin  {equation}
\frac {d  E}  {dt }
=
\frac {E }  {\tau_{II} }   \quad,
\end {equation}
where $\tau_{II}$ is the  typical time-scale,
\begin {equation}
\frac{1}{\tau_{II}}  = \frac {4} {3 }
( \frac {u^2} {c^2 }) (\frac {c } {L_{II} })
\quad ,
\label {tau2}
\end   {equation}
where $u$ is the velocity of the accelerating cloud
belonging  the advancing shell of the SB,
$c$  is the velocity of light,
and $L_{II}$  is the
mean free path between clouds \citep[][Eq. 4.439]{lang}.
The mean free path between the accelerating clouds
in the Fermi II mechanism can be found from the following
inequality in time:
\begin{equation}
\tau_{II} < \tau_{sync}
\quad  ,
\end{equation}
which  corresponds to  the following  inequality for the
mean free  path  between scatterers
\begin{equation}
L  <
\frac
{
1.72\,10^5 \,{u}^{2}
}
{
H\sqrt {H\nu}{c}^{2}
}
\,\mathrm{pc}
\quad  .
\end{equation}
The mean free path length for an SB 
at  [S~II] continuum,
which means 7040 $\AA$ or 704  nm or
$4.258\,10^{14}$\ Hz,  
gives
\begin{equation}
L <
\frac
{
2.94\,10^{-6}\,{{\it u_1}}^{2}
}
{
{{\it H_5}}^{3/2}
}
\, \mathrm{pc}
\end{equation}
where  
$u_{1}$ is the   velocity of the accelerating  
cloud  
expressed in
km/s,
and  $H_5 =$  is the  magnetic field expressed
in   units of $10^{-5}$ Gauss.
When this inequality  is verified, the direct
conversion of the rate   of  kinetic energy
into radiation can be adopted.
Recall that the Fermi II  mechanism  produces
an  inverse power law
spectrum in the
energy
of the type
$
N (E) \propto  E ^{-\gamma}
$
or an  inverse power law  in the observed frequencies
$
N (\nu) \propto   \nu^{\beta }
$
with  $ \beta= - \frac{\gamma -1}{2}$
\citep{lang,Zaninetti2011a}.
The strong shock accelerating mechanism  
named  Fermi I
was  introduced by  
\cite{Bell_I}
and 
\cite{Bell_II}.
The energy  gain  relative to a particle that is crossing
the shock is:
\begin  {equation}
\frac {d  E}  {dt }
=
\frac {E }  {\tau_{I} }   \quad,
\end {equation}
where $\tau_{I}$ is the  typical time-scale,
\begin {equation}
\frac{1}{\tau_{I}}  = \frac {2} {3 }
( \frac {u} {c }) (\frac {c } {L_{I} })
\quad ,
\end   {equation}
where $u$ is the velocity of the shock of the SB,
$c$  is the velocity of light,
and $L_{I}$  is the
mean free path between scatterers.
This process produces an energy spectrum of
electrons
of the type:
\begin {equation}
N (E) dE \propto E^{-2} dE
\quad,
\end   {equation}
see  \cite{longair}.
The  two mechanisms  ,
Fermi I and Fermi II, 
produce the  same results  when
\begin{equation}
\frac{L_{II}} {L_{I}} =
2 \frac{u}{c}
\quad .
\end{equation}

\label{lastpage}

\end{document}